\documentclass[runningheads]{llncs}

\usepackage{amsfonts, mathtools}
\usepackage{microtype}                                                                              \usepackage[linesnumbered, ruled, vlined, procnumbered, nokwfunc]{algorithm2e}                      \usepackage{booktabs, tabularx}                                                                     \usepackage{environ}                                                                                \usepackage{soul, color}                                                                            \usepackage[dvipsnames, table]{xcolor}                                                              \usepackage{pifont}                                                                                 \usepackage[normalem]{ulem}                                                                         \usepackage{float}                                                                                  \usepackage{adjustbox}                                                                              \usepackage{pdflscape}                                                                              \usepackage{textgreek}                                                                              \usepackage{hyperref}                                                                               
\hypersetup{
colorlinks = true,
  linktoc = all,
  citecolor = RoyalBlue!95!Black,
  linkcolor = RedOrange!85!Black,
  urlcolor = PineGreen!95!Black,
  filecolor = cyan, 
  bookmarksdepth = 2
}

\newcommand*{\prob}[1]{\textsc{#1}}                                                                 \newcommand*{\probc}[1]{\textsf{#1}}                                                                \newcommand*{\algo}[1]{\texttt{#1}}                                                                 \newcommand*{\algoi}[2]{\texttt{#1}(#2)}                                                            \SetKw{Read}{read}
\SetKw{Write}{write}
\SetKw{Break}{break}
\SetKw{Continue}{continue}
\SetAlgoCaptionSeparator{,}

\DeclarePairedDelimiter{\brnX}{(}{)}

\DeclarePairedDelimiter{\brcX}{\{}{\}}
\DeclarePairedDelimiter{\absX}{\lvert}{\rvert}

\newcommand*{\brn}{\brnX*}                                                                                                                                                                                                                              \newcommand*{\brc}{\brcX*}                                                                          \newcommand*{\abs}{\absX*}

\DeclareMathOperator{\ohop}{O}                             
                             
\DeclareMathOperator{\omop}{\Omega}

\newcommand*{\Oh}[1]{\ohop\brn{#1}}                                                                                                                                 \newcommand*{\Om}[1]{\omop\brn{#1}}

\DeclareMathOperator{\Vop}{V}

\DeclareMathOperator{\Nbop}{N}
\newcommand*{\V}[1]{\Vop\brn{#1}}                                                                                                                                                                                                       \newcommand*{\Nbr}[2]{\Nbop_{#1}\brn{#2}}                                                           \let\degd\deg
\let\deg\relax
\newcommand*{\deg}[1]{\degd\brn{#1}}                                                                                                                          
\newcommand*{\N}{\mathbb{N}}                                                                                                                                                                                                                                                                                                                                                                                                                                                                                                          \newcommand*{\setb}[2]{\left\{#1 \mid #2\right\}}

\newcommand*{\dnimply}{\kern.6em\not\kern-.6em \implies}                                                                                                                                                                              \newcommand*{\YES}{\texttt{YES}}                                                                    \newcommand*{\NO}{\texttt{NO}}                                                                      

\newcommand*{\sgn}[1]{\sgn\brn{#1}}                                                                                                                                     \newcommand*{\oper}[1]{\operatorname{#1}}                                                                                                                                \newcommand*{\exS}[1]{\textsuperscript{#1}}                                                                                                                                                                                                                                                              
\makeatletter
\newcommand{\ptitle}[1]{\gdef\prob@title{#1}}                                                       \newcommand{\pobject}[1]{\gdef\prob@object{#1}}                                                     \newcommand{\pquery}[1]{\gdef\prob@query{#1}}                                                       \newcommand{\pparam}[1]{\gdef\prob@param{#1}}                                                       
\NewEnviron{prblm}[1][]{\def\prob@type{#1}\def\prob@topt{opt}\def\prob@title{}\def\prob@object{}\def\prob@query{}\def\prob@param{}\ifx\prob@type\prob@topt \def\prob@qword{Solution}\else \def\prob@qword{Question}\fi \BODY
  \ifx\prob@param\@empty
    \def\prob@trow{{\large\prob@title}}\else \def\prob@trow{\begin{tabular*}{\textwidth}{@{\extracolsep{\fill}}lr}{\large\prob@title} & \textbf{Parameter:}~\prob@param \end{tabular*}}\fi \par\noindent\ignorespaces \ignorespacesafterend \\*[-1ex]\fbox{\hspace{0.5em}\begin{minipage}{0.95\textwidth}\vspace{1.2ex}
      \prob@trow\\[2\fboxsep]
      \textbf{Instance:}~\prob@object\\
      \textbf{\prob@qword:}~\prob@query \vspace{1.4ex}
    \end{minipage}\hspace{0.5em}}\par }
\makeatother

\newcommand*{\cFPT}{\probc{FPT}}

\newcommand*{\cLAdv}{\probc{LOGSPACE + advice}}
\newcommand*{\cSlL}{\probc{SLICEWISE LOGSPACE}}
\newcommand*{\cL}{\probc{L}}
\newcommand*{\cNL}{\probc{NL}}

\newcommand*{\cParaL}{\probc{Para-L}}

\newcommand*{\cP}{\probc{P}}

\newcommand*{\cXL}{\probc{XL}}
\newcommand*{\cFPTXL}{\probc{FPT + XL}}

\newcommand*{\pChVD}{\prob{Chordal Vertex Deletion}}

\newcommand*{\pCVD}{\prob{Cluster Vertex Deletion}}
\newcommand*{\pdCol}{d\text{--}\prob{Colouring}}
\newcommand*{\pDelPi}{\prob{Del\textPi}}
\newcommand*{\pDFVS}{\prob{Directed Feedback Vertex Set}}
\newcommand*{\pHS}{\prob{Hitting Set}}
\newcommand*{\pdHS}{d\text{--}\prob{Hitting Set}}

\newcommand*{\pDLF}{\prob{Deletion to Linear Forests}}
\newcommand*{\pDPo}{\prob{Deletion to Pathwidth 1}}

\newcommand*{\pqPartDisjPiFree}{q\text{--}\prob{PartDisj\textPi{}Free}}

\newcommand*{\pFVS}{\prob{Feedback Vertex Set}}

\newcommand*{\pLongCycle}{\prob{Longest Cycle}}
\newcommand*{\pIS}{\prob{Independent Set}}
\newcommand*{\pClq}{\prob{Clique}}
\newcommand*{\pLongPath}{\prob{Longest Path}}
\newcommand*{\pIndPiFDS}{\prob{Ind\textPi{}FreeDel}}

\newcommand*{\pLIndPiS}{\prob{LargeInd\textPi{}Subgraph}}

\newcommand*{\pMOdS}{\prob{Min-Ones}\ d\text{--}\prob{SAT}}

\newcommand*{\pOCT}{\prob{Odd Cycle Transversal}}

\newcommand*{\pPlan}{\prob{Planarization}}

\newcommand*{\pSVD}{\prob{Split Vertex Deletion}}

\newcommand*{\pThVD}{\prob{Threshold Vertex Deletion}}
\newcommand*{\pTW}{\prob{Treewidth}}

\newcommand*{\pVC}{\prob{Vertex Cover}}
 
\title{Space-Efficient FPT Algorithms}
\author{Arindam Biswas\inst{1} \and Venkatesh Raman\inst{1} \and Srinivasa Rao Satti\inst{2} \and Saket Saurabh\inst{3}}
\authorrunning{A.\ Biswas et al.}
\institute{The Institute of Mathematical Sciences, HBNI, Chennai, India\\\email{\{barindam,vraman,saket\}@imsc.res.in} \and Norwegian University of Science and Technology, Trondheim, Norway\\\email{srinivasa.r.satti@ntnu.no} \and University of Bergen, Bergen, Norway}

\begin{document}
\maketitle              \begin{abstract}
We prove algorithmic results showing that a number of natural parameterized problems are in the restricted-space parameterized classes $\cParaL$ and $\cFPTXL$. The first class comprises problems solvable in $f(k)\, n^{\Oh{1}}$ time using $g(k) + \Oh{\log{n}}$ bits of space ($k$ is the parameter and $n$ is the input size; $f$ and $g$ are computable functions). The second class comprises problems solvable under the same time bound, but using $g(k) \log{n}$ bits of space instead.

Earlier work on these classes has focused largely on their structural aspects and their relationships with various other classes. We complement this with $\cParaL$ and $\cFPTXL$ algorithms for a restriction of $\pHS$, some graph deletion problems where the target class has an infinite forbidden set characterization, a number of problems parameterized by vertex cover number, and $\pFVS$.

\end{abstract}

\section{Introduction}
We consider the task of identifying fixed-parameter tractable (FPT) problems that admit space-efficient algorithms. Towards this end, we devise algorithms for various parameterized problems that run in time $f(k)\, n^{\Oh{1}}$ and use either $g(k) + \log{n}$ or $g(k) \log{n}$ bits of space, where $n$ denotes the input size, $k$ denotes the parameter, and $f, g: \N \to \N$ are computable functions.

\subsubsection{Previous Work} Work on restricted-space classes of parameterized problems was initiated by Cai et al.~\cite{CCDF1997AnnPureApplLogic}, who defined the classes $\cLAdv$ and $\cSlL$, which we call $\cParaL$ and $\cXL$ respectively. Among other things, they showed that $\pVC$ under the usual solution-size parameterization is in $\cParaL$. Continuing this line of work, Flum and Grohe~\cite{FG2003InfComput} showed that the parameterized model-checking problem of first-order formulas on graphs of bounded degree is in $\cParaL$. As a consequence, many standard parameterized graph problems are in $\cParaL$ when restricted to bounded-degree graphs. Then, Elberfeld et al.~\cite{EST2012IPEC} introduced two new restricted-space classes and gave completeness results for those classes, showing that there are problems fixed-parameter tractable (FPT) problems which are outside $\cParaL$ under the assumption that $\cL{} \neq \cNL$. In particular, they identified $\pDFVS$ as one such problem. In contrast, $\pDFVS$ restricted to tournaments can be shown to be in $\cParaL$ (Corollary~\ref{corr:del_dhs}). Later work in this setting includes the paper of Chen and M\"{u}ller~\cite{CM2015TOCT}, who studied additional complexity-theoretic aspects of restricted-space classes and showed that $\pLongPath$ is in $\cParaL$.

$\pdHS$, a generalization of $\pVC$ to hypergraphs was shown by Fafanie and Kratsch~\cite{FK2015MFCS} to be kernelizable in logarithmic space, which puts it in $\cParaL$ (see Lemma~\ref{lemm:para-l_kern}). As a consequence they showed that various graph deletion problems where the target classes are characterized by finite forbidden sets are also in $\cParaL$. For related results, see~\cite{BT2020TheoryComputSyst,BST2020SWAT} which give constant-time parallel kernelization algorithms for $\pHS$.

\subsubsection{Results, Techniques and Organization} Section~\ref{sect:deletion} begins by showing that $\pHS$ (with possibly unbounded set sizes) can be placed in $\cParaL$ under the restriction that pairwise intersections of sets in the instances have bounded sizes.

Then in Section~\ref{sect:deletion_problems}, we extend the results of Fafianie and Kratsch~\cite{FK2015MFCS} by showing that some graph deletion problems where the target classes have infinite forbidden sets are also in $\cParaL$.
As applications, we show that $\pDLF$ and $\pDPo$ are in $\cParaL$.

Initially, studies in parameterized complexity concerned the solution-size parameter or width parameters such as treewidth and cliquewidth. In recent times, vertex cover number as a parameter has become a subject of serious investigation~\cite{Jan2013thesis,FJP2014JCSS,Jd2020SWAT}. In Section~\ref{sect:vc_parameterization} we give $\cParaL$ algorithms for problem parameterized by vertex cover number. We devise a logarithmic-space variant of a general kernelization algorithm~\cite{FJP2014JCSS} for problems parameterized by vertex cover number, and as a consequence obtain $\cParaL$ algorithms for $\pFVS$, $\pOCT$, $\pChVD$, $\pPlan$, $\pLongPath$, $\pLongCycle$ and $\pdCol$ under this parameterization.

Finally, in Section~\ref{sect:fvs}, we address problems that are not known to be in $\cParaL$, but are known to be in $\cFPT$ as well as in $\cXL$, i.e.\ solvable using $f(k) + \Oh{\log n}$ bits of space. We ask whether there are algorithms for those problems that \emph{simultaneously} run in time $f(k)\, n^{\Oh{1}}$ and in $f(k) \log{n}$ bits of space. Towards this end, we devise such an algorithm for $\pFVS$. This is achieved via a careful space-efficient implementation of the iterative-compression algorithm of Chen et al.~\cite{CFL+2008JCSS}.

 \section{Preliminaries}
\subsubsection{Notation and Definitions}
For a graph $G$, we denote its vertex set by $V(G)$ and edge set by $E(G)$.
A class of graphs $\Pi$ is said to be \emph{characterized} by a set $\Phi$ of induced subgraphs if $\Pi$ consists of precisely those graphs that do not include induced subgraphs isomorphic to any graph in $\Phi$.

\subsubsection{Appendix} Details for all items marked $\dagger$ can be found in the Appendix.

\subsubsection{Parameterized Problems and Space Classes}
Let $A$ be a decision problem over the alphabet $\Sigma$ and $t: \Sigma^* \to \N$ be a computable function. The pair $(A, t)$ is called a \emph{parameterized} problem and $t$ is called the \emph{parameterization}.

An instance of the problem is a pair $(I, t(I))$, where $I$ is an instance of $A$. The problem $(A, t)$ is said to be \emph{fixed-parameter tractable} (FPT) if there is an algorithm which solves any instance $(I, t(I))$ in time $f(t(I)) {\abs{I}}^c$, where $f: \N \to \N$ is a computable function, $n = \abs{I}$, and $c > 0$ is a constant. Later on, we refer to such running times as \emph{FPT time}.

The problem is \emph{kernelizable} if there is an algorithm which in time polynomial in $n$ computes an instance $I'$ such that $\abs{I'} = g(t(I))$ for some computable function $g: \N \to \N$ and $I$ is a \YES{} instance if and only if $I'$ is a \YES{} instance. The instance $I'$ is called a \emph{kernel}. Kernels are obtained through what are called \emph{reduction rules}, which transform a given instance into another. And the rule is \emph{safe} if the two instances are equivalent.

In this paper, we focus on the \emph{space complexity} aspect of parameterized problems. A natural problem class to consider in this setting is $\cParaL$, defined below.

\begin{definition}[$\cParaL$; Cai et al.~\cite{CCDF1997AnnPureApplLogic}]
	$\cParaL$ is the class of all parameterized problems $(A, t)$ for which there is a deterministic algorithm which solves instances $(I, t(I))$ using space $f(k) + \Oh{\log{n}}$, where $f: \N \to \N$ is a computable function, $k = t(I)$ and $n = \abs{I}$.
\end{definition}

We also consider an analogous class $\cXL$, where the space bound is $f(k) \cdot \log{n}$ instead of $f(k) + \Oh{\log{n}}$. The two classes are known to be distinct unless $\cP{} = \cL$ (Theorem 3.1 of~\cite{CCDF1997AnnPureApplLogic}).
\begin{definition}[$\cXL$; Cai et al.~\cite{CCDF1997AnnPureApplLogic}]
	$\cXL$ is the class of all parameterized problems $(A, t)$ for which there is a deterministic algorithm which solves instances $(I, t(I))$ using space $f(k) \cdot \log{n}$, where $f: \N \to \N$ is a computable function, $k = t(I)$ and $n = \abs{I}$.
\end{definition}

\subsubsection{Machine Model}
We use the standard model for space-efficient algorithms. The input to an algorithm resides in read-only memory and can be randomly accessed, while the output is written to a stream which cannot be read from. The algorithm uses a small number of read-write cells as auxiliary memory, where each cell can hold a word of size $\Oh{\log{n}}$ bits ($n$ is the input size). The machine is a unit-cost RAM in which basic arithmetic and logic operations involving two words take constant time.

A basic result in parameterized complexity theory is that a problem is fixed-parameter tractable if and only if it is kernelizable. The following lemma is an easy consequence of the result.
\begin{lemma}\label{naivespace}
If a parameterized problem is fixed-parameter tractable, then it can be solved using $f(k) + n^{O(1)}$ space where $f$ is some function of the parameter $k$, and $n$ is the input size.
\end{lemma}

The next result is implicit in~\cite{CCDF1997AnnPureApplLogic}; we include it here for the sake of clarity.
\begin{lemma}\label{lemm:para-l_kern}\exS{$\dagger$}
	Let $(A, t)$ be a parameterized problem. For any computable function $f: \N \to \N$, the following statements are true.
	\begin{itemize}
		\item If $(A, t)$ can be solved in space $f(k) + \Oh{\log{n}}$, then it can be kernelized in space $\Oh{\log{n}}$.
		\item If $(A, t)$ can be kernelized in space $f(k) + \Oh{\log{n}}$, then it can be solved in space $g(k) + \Oh{\log{n}}$, where $g: \N \to \N$ is a computable function.
	\end{itemize}
\end{lemma}

 \section{Hitting Set and Graph Deletion Problems}\label{sect:deletion}
We begin by examining two known small-space kernelization results. The first is a combination a logarithmic-space implementation of the Buss rule~\cite{BG1993SICOMP} for $\pVC$ and the observation that the kernel produced is itself a vertex cover.

\begin{proposition}[Cai et al.~\cite{CCDF1997AnnPureApplLogic}, Theorem 2.3]\label{prop:vc_para-l}
	There is an algorithm which takes as input a graph $G$ and $k \in \N$, and using space $\Oh{\log n}$, either answers correctly that $G$ has no vertex cover of size at most $k$ or produces a vertex cover of size at most $2k^2$.
\end{proposition}

The next result gives an equivalent algorithm for $\pdHS$.
\begin{proposition}[Fafianie and Kratsch~\cite{FK2015MFCS}, Theorem 1]\label{prop:dhs_log_kern}
	There is an algorithm which takes as input an instance of $\pdHS$, i.e. a universe $U$, a family $\mathcal{F}$ of subsets of $U$ of size $d$ each and an integer $k$, and in time $n^{\Oh{d^2}}$ and $\Oh{d^2 \log{n}}$ bits of space, either answers correctly that the instance has no hitting set of size at most $k$ or produces an instance $(U', \mathcal{F}', k)$ such that
	\begin{itemize}
	 	\item $U' \subseteq U,\ \abs{U'} \leq d{(k + 1)}^d$,
	 	\item $\mathcal{F}' \subseteq \mathcal{F},\ \abs{\mathcal{F}'} \leq {(k + 1)}^d$, and
	 	\item for all $S \subseteq U$ with $\abs{S} \leq k$, $S\ \text{is a minimal hitting set for}\ \mathcal{F}$ if and only if $S\ \text{is a minimal hitting set for}\ \mathcal{F}'$.
	 \end{itemize}
\end{proposition}

Since the above algorithm kernelizes $\pdHS$ in logarithmic space (for constant $d$), we have the following result as a consequence of Lemma~\ref{lemm:para-l_kern}.

\begin{corollary}\label{corr:dhs_para-l}
	For constant $d$, $\pdHS$ is in $\cParaL$.
\end{corollary}

\subsection{Deletion Problems}\label{sect:deletion_problems}
We begin this section by showing that if a certain restriction of a general graph deletion problem ($\pDelPi$, defined below) is in $\cParaL$, then the entire problem is in $\cParaL$. Combining this with a method for partially encoding these problems as $\pHS$, we show that $\pDLF$ and $\pDPo$ are in $\cParaL$. The problem $\pDelPi$ has the following form.

\textbf{Instance:} $(G, k)$, where $G$ is a graph and $k \in \N$\\
\hspace*{\parindent}\textbf{Question:} Is there a set $S \subseteq V$ with $\abs{S} \leq k$ such that $G - S \in \Pi$?\\

When $\Pi$ is characterized by a finite set of forbidden induced subgraphs $\Phi$, instances of $\pDelPi$ can be modelled as $\pdHS$. As an example, consider $\pCVD$, where the objective is to delete up to $k$ vertices in an input graph such that the resulting graph is a cluster graph. The class of cluster graphs is exactly the class of graphs which do not have a path on three vertices $P_3$ as induced subgraphs. Because of this, an instance $(G, k)$ of the problem can be encoded as an instance $(U, \mathcal{F}, k)$ of $3$--$\pHS$, where $U = \V{G}$ and $\mathcal{F} = \setb{\V{P}}{P\ \text{is an induced}\ P_3\ \text{in}\ G}$. The family of sets $\mathcal{F}$ is can be generated in logarithmic space: enumerate all subsets of $\V{G}$ of size $3$ and output those that induce a $P_3$. t is easy to see that $(U, \mathcal{F}, k)$ is a \YES{} instance of $3$--$\pHS$ if and only if $(G, k)$ is a \YES{} instance of $\pCVD$. This family can now be used as input to a $3$--$\pHS$ algorithm, and because of Corollary~\ref{corr:dhs_para-l}, the instance can be solved in $f(k) + \log{n}$ bits of space. A variety of other problems can be solved in a similar fashion, leading to the following result.

\begin{corollary}\label{corr:del_dhs}
	$\pCVD$, $\pDFVS$ restricted to tournaments, $\pSVD$ and $\pThVD$ are in $\cParaL$.
\end{corollary}

In what follows, we describe the main result of this section, which extends Corollary~\ref{corr:del_dhs} to also include some variants of $\pDelPi$ where $\Pi$ is characterized by an infinite set of forbidden induced subgraphs.

\begin{theorem}\label{thm:restr_para_l}\exS{$\dagger$}
	Let $\Psi \supseteq \Pi$ be a class of graphs and let $\prob{ResDel\textPi}$ be the restriction of $\pDelPi$ to graphs in $\Psi$. If $\Psi$ is characterized by a finite set $\Phi$ of forbidden induced subgraphs and $\prob{ResDel\textPi} \in \cParaL$, then $\pDelPi \in \cParaL$.
\end{theorem}

Since $\pDelPi$ restricted to $\Psi$ is in $\cParaL$, there is an algorithm which solves restricted instances $(H, k)$ of $\pDelPi$ using space $f(k) + \Oh{\log{n}}$ ($f: \N \to \N$, a computable function). Let \algo{SolveDel\textPi{}On\textPsi{}} be such an algorithm. Using this as a subroutine, Algorithm~\ref{algo:SolveDelPi} solves $\pDelPi$.

\begin{algorithm}[h]
\KwIn{$(G, k)$, where $G$ is a graph and $k \in \N$}
\KwOut{\YES{} if there is a set $S \subseteq V$ of size at most $k$ such that $G - S \in \Pi$, and \NO{} otherwise}

	let $c_1, \dotsc, c_t$ be the sizes of the sets in $\Phi$\;
	$d \gets \max \brc{c_1, \dotsc, c_t},\ V \gets \V{G}$\;
	$\mathcal{F} \gets \setb{S \subseteq V}{G[S]\ \text{is isomorphic to some}\ H \in \Phi}$\;
	$(V', \mathcal{F}', k) \gets \algoi{KernelizeDeeHS}{d, V, \mathcal{F}, k}$\;
	\Return{\algoi{BranchAndCall}{$G, \mathcal{F}', \emptyset, k$}}\;
	\SetKwFunction{bncproc}{BranchAndCall}
  \SetKwProg{subproc}{Procedure}{}{}
  \subproc{\bncproc{$G, \mathcal{F}', S, l$}}{
  	\If{$l < 0$}{
  		\Return{\NO{}}\;
  	}
  	\eIf(\tcp*[f]{$G - S$ is in $\Psi$}){$S$ hits all of $\mathcal{F}'$}{
  		\If{\algoi{SolveDel\textPi{}On\textPsi{}}{$G - S, l$} returns \YES{}}{
  			\Return{\YES{}}\;
			}
  	}{
  		find a set $A \in \mathcal{F}'$ such that $A \cap S = \emptyset$\;
  		\For{$v \in A$}{
  			\If{\algoi{BranchAndCall}{$G, \mathcal{F}', S \cup \brc{v}, l - 1$} returns \YES{}}{
  				\Return{\YES{}}\;
  			}
  		}
  	}
  	\Return{\NO{}}\;
  }

\caption{SolveDel\textPi: solve \pDelPi{} with access to \algo{KernelizeDeeHS}, which kernelizes \pdHS{}, and \algo{SolveDel\textPi{}on\textPsi{}}, which solves the restriction of \pDelPi{} to \textPsi{}}\label{algo:SolveDelPi}
\end{algorithm}

\begin{lemma}\label{lem:correctness}\exS{$\dagger$}
	Algorithm~\ref{algo:SolveDelPi} solves $\pDelPi$ in $g(k) + \Oh{\log{n}}$ bits of space ($g: \N \to \N$, a computable function).
\end{lemma}

The claim of Theorem~\ref{thm:restr_para_l} now readily follows from the preceding discussion. The next result is an application of Theorem~\ref{thm:restr_para_l} which makes use of \algo{SolveDel\textPi{}On\textPsi{}} routines for some specific graph classes $\Psi$. It is pertinent here to note that the target classes $\Pi$ do not have characterizations by finite forbidden sets. 

\begin{corollary}\label{corr:dlf_pw1}\exS{$\dagger$}
	\pDLF{}  and \pDPo{} are in $\cParaL$.
\end{corollary}

Both problems are related to $\pFVS$ (delete vertices so that the resulting graph is a forest): a solution to either problem is also a feedback vertex set. While we are able to show that \pDLF{} and \pDPo{} are in \cParaL{}, the known kernelization algorithms and fixed-parameter algorithms for $\pFVS$ (where the parameter, i.e.\ the solution size, is smaller) use strategies which appear difficult to carry out in the $\cParaL$ setting.

\subsection{$\pHS$ with Bounded Intersection}
Proposition~\ref{prop:dhs_log_kern} shows that one can, in logarithmic space, compute kernels for instances of $\pdHS$ in which each set is of size $d$. In this section, we obtain a similar result for the scenario where not the set sizes, but the intersection between any pair of sets is bounded by a constant $s$. Interestingly, the kernelization question for this natural $\pHS$ variant has not been addressed in earlier works, even with no constraints on space.

\subsection*{Case $s = 1$}
We first consider the case $s = 1$, i.e.\ when the instances are linear hypergraphs~\cite{Ber1989book}. Let $(U, \mathcal{F}, k)$ be an instance of $\pHS$ where the intersection between any pair of sets is of size at most $s = 1$. The algorithm consists of the following sequence of reduction rules.

\begin{description}
	\item[Rule $1$] Let $S$ be the set of all elements that appear in at least $k + 1$ sets in $\mathcal{F}$. If $\abs{S} > k$, return \NO{}. Otherwise, delete all sets in $\mathcal{F}$ that intersect with $S$ and delete all elements of $S$ from $U$. Set $k \gets k - \abs{S}$.
	
	\item[Rule $2$] If the number of sets remaining is more than $k^2$, then return NO.

	\item[Rule $3$] For each remaining set, determine the elements that appear in no other set. Delete all but one of them from the set and from the universe.
\end{description}

The following lemma establishes the correctness of the above rules and gives a bound on the size of the final instance.
\begin{lemma}\label{lemm:hs_corr}\exS{$\dagger$}
	Reduction Rules $1$, $2$ and $3$ produce a kernel $(U', \mathcal{F}', k')$, where $\abs{U'} \leq k^2$ and $\abs{\mathcal{F}'} \leq k^4$.
\end{lemma}

\subsubsection{Space-Efficient Implementation}
Reduction Rule $1$ can be implemented using two counters in addition to a constant number of counters for iteration. The first counter counts, for each element, the number of sets it appears in. The second counter counts the number of elements for which the first counter is larger than $k$. If the second counter goes beyond $k$, we return \NO{}. Otherwise, we set $k \gets k - k'$, where $k'$ is the value of the second counter.

To implement Reduction Rule $2$, we set up two additional counters and run through all the sets. The first counter checks for each set whether there is any element in it that appears more than $k$ times in $\mathcal{F}$. If no element of the set appears in more than $k$ times, then the set survives after Reduction Rule $1$, and we increment the second counter. Once all sets have been processed in this manner, if the value of the second counter is more than $k^2$, we return \NO{}.

To implement Reduction Rule $3$, we first determine as for Rule $2$ which sets $S$ survive the application of Rule $1$. We suppress the sets that do not survive, and output a subset of $S$ as follows. Observe that at this point, elements in each surviving set appear at most $k$ times in the entire instance. For each element of $S$, we count the number of times it appears in the instance. We output all elements that appear more than once, and of the ones that appear exactly once, we output the first such element in the set. The latter step can be performed using the following rule for each element $x$ in $S$ that appears exactly once in the instance: output $x$ only if there is no other element in $S$ before $x$ which also appears exactly once in the instance. This only uses a constant number of additional counters.

Observe that in all the steps above, only a constant number of additional counters are used, from which it readily follows that the entire procedure uses $\Oh{\log{n}}$ bits of space. Combining this with Lemma~\ref{lem:correctness}, we have the following result.

\begin{theorem}
    Given a number $k \in \N$ and a family of sets from a finite universe of size $n$ where each pair of sets intersect in at most one element, using $\Oh{\log{n}}$ bits of space, one can either determine that the given instance is a \NO{} instance of $\pHS$ or output an equivalent instance consisting of a family of at most $k^2$ sets with at most $k^2$ elements each such that each pair of sets intersects in at most one element.
\end{theorem}

Combining the above result with Lemma~\ref{lemm:para-l_kern}, we have the following corollary.
\begin{corollary}
	$\pHS$ restricted to instances where any pair of sets intersect in at most one element is in $\cParaL$.
\end{corollary}

\subsection{Case $s > 1$}\label{ssct:bd_hs_dg1}
In this case, the instance can be reduced using similar rules as before, with a sequence of $s - 1$ rules replacing Reduction Rule $1$ in the $s = 1$ case.

\begin{description}
	\item[Rule $i$ ($i = 1, \dotsc, s - 1$)] If there is an $(s+1-i)$-element subset of the universe that appears as a subset of $k^i+1$ sets, then replace all those sets with just the $(s+1-i)$-element set.

	For example, Reduction Rule $1$ ensures that if an $s$-element subset of the universe appears in $k+1$ sets, then all those sets are replaced with just the common $s$-element subset.
	
	\item[Rule $s$] Let $S$ be the set of all elements that appear in at least $k^s + 1$ sets in $\mathcal{F}$. If $\abs{S} > k$, return \NO{}. Otherwise, delete all sets in $\mathcal{F}$ that intersect with $S$ and delete all elements of $S$ from $U$. Set $k \gets k - \abs{S}$.

	\item[Rule $(s+1)$] If the number of sets remaining is more than $k^{s + 1} + 1$, then return \NO{}.

	\item[Rule $(s+2)$] For each remaining set, determine the elements that appear in no other set. Delete all but one of them from the set and from the universe.
\end{description}

As in the $s = 1$ case, the above rules ensure that the size of the final instance is bounded by a function of $k$. In addition, the rules can be applied using $\Oh{s \log{n}}$ bits of space per rule. Owing to space constraints, we defer the space-efficient implementation of the above rules to Appendix~\ref{apdx:deletion} and state our result directly.

\begin{theorem}\exS{$\dagger$}
    Given a family of sets from an $n$-element universe where each pair of sets intersects in at most $s$ elements, using $\Oh{s^2 \log{n}}$ bits of space, one can either determine that the given instance is a \NO{} instance of $\pHS$ or output an equivalent family of at most $k^{s + 1}$ sets with at most $s k^{s + 1}$ elements each such that each pair of sets intersects in at most $s$ elements.
\end{theorem}

Combining the above result with Lemma~\ref{lemm:para-l_kern} yields the following corollary.

\begin{corollary}
	For constant $s$, $\pHS$ restricted to instances where any pair of sets intersect in at most $s$ elements is in $\cParaL$.
\end{corollary} \section{Vertex Cover Number as Parameter}\label{sect:vc_parameterization}
With respect to the standard parameter (solution size), problems such as $\pTW$ and $\pFVS$ are not known to be in $\cParaL$. In this section, we show that when parameterized by vertex cover number, these problems are in fact in $\cParaL$.

Elberfeld et al.~\cite{EJT2010FOCS} give an algorithm for computing the treewidth of an undirected graph using space $f(k) \cdot \log{n}$, which puts the problem in $\cXL$, but it is not known if the problem is in $\cParaL$. For $\pFVS$, the obvious ``guess and verify'' algorithm puts it in $\cXL$, but does not run in FPT time. The known fixed-parameter algorithms for these problems (see e.g.~\cite{CFL+2008JCSS,RSS2006TALG}) involve complex branching strategies and it seems unlikely that they can be carried out using space $f(k) + \Oh{\log{n}}$.

The following definition captures classes of graphs which are closed under  certain modification operations. Such graph classes encompass ``target'' graphs for many graph modification and partitioning problems, and can be used to show the kernelizability of these problems, as shown later.

\begin{definition}[Bounded-Adjacency Characterization; Fomin et al.~\cite{FJP2014JCSS}]
	Let $\Pi$ be a class of graphs and $c_{\Pi} \in \N$ be a constant. If for any graph $G \in \Pi$ and any vertex $v \in \V{G}$, there is a set of vertices $D \subseteq \V{G} \setminus \brc{v}$ with $\abs{D} \leq c_{\Pi}$ such that adding or removing any number of edges between $v$ and $\V{G} \setminus{D}$ produces a graph which is also in $\Pi$, then $\Pi$ is said to be characterized by $c_{\Pi}$ adjacencies. 
\end{definition}

Consider the following problem, defined with respect to a graph class $\Pi$ characterized by $c_{\Pi}$ adjacencies.
\begin{description}
	\item[$\pIndPiFDS$]\hfill\\
		$\pi$ is a class of \emph{non-empty} graphs and there is a non-decreasing polynomial $p$ such that every vertex minimal graph $G \in \Pi$ satisfies $\abs{G} \leq p(\tau)$, where $\tau$ is the size of a minimum vertex cover for $G$.\\
		\textbf{Instance:} $(G, k)$, where $G$ is a graph and $k \in \N$\\
		\textbf{Question:} Is there a set $S \subseteq V$ with $\abs{S} \leq k$ such that $G - S$ includes no graph in $\Pi$ as an induced subgraph?
\end{description}

Fomin et al.~\cite{FJP2014JCSS} give a generic procedure \algo{Reduce} (see Appendix~\ref{apdx:vc_param}), which kernelizes the problem as described in the following proposition.  

\begin{proposition}[Theorem 2, Fomin et al.~\cite{FJP2014JCSS}]\label{prop:reduce_pvc_kernels}
	For an instance $(G, k)$ of $\pIndPiFDS$ with a vertex cover $X$ of size $r \geq k$, \algoi{Reduce}{$G, X, r + p(r), c_{\Pi}$} produces a kernel of size ${r}^{c_{\Pi}} (r + p(r))$.
\end{proposition}

We now state the main result of this section, which is obtained by carefully combining the logarithmic-space $\pVC$ kernelization from earlier with a modified \algo{Reduce} procedure.
\begin{theorem}\label{thm:kern_pvc}
	$\pIndPiFDS$ is in $\cParaL$.
\end{theorem}

While the \algo{Reduce} procedure of Fomin et al.~\cite{FJP2014JCSS} can be used to kernelize $\pIndPiFDS$, it is not clear that the procedure can be carried out in small space. We use a modified reduction procedure \algo{ReduceLog} (see Appendix~\ref{apdx:vc_param}) instead, which uses space $\Oh{\log{n}}$, where $n$ is the order of the input graph.

\begin{lemma}\label{lemm:reduce_log_pvc}\exS{$\dagger$}
	For an instance $(G, k)$ of $\pIndPiFDS$ with a vertex cover $X$ of size $r \geq k$, \algoi{ReduceLog}{$G, X, r + p(r), c_{\Pi}$} produces a kernel of size ${r}^{c_{\Pi}} (r + p(r))$. The procedure uses $\Oh{\log{n}}$ bits of space, where $n = \abs{G}$.
\end{lemma}

The proof of Theorem~\ref{thm:kern_pvc} is now straightforward.

\begin{proof}[Theorem~\ref{thm:kern_pvc}]
	Lemma~\ref{lemm:reduce_log_pvc} shows that given access to the input instance $(G, k)$ and a vertex cover $X$ for $G$, $\pIndPiFDS$ is kernelizable (using \algo{ReduceLog}) in space $\Oh{\log{n}}$, where $n = \abs{G}$. For a solution-size value $k$ (the size of the vertex cover), the $\pVC$ kernelization of Proposition~\ref{prop:vc_para-l} either correctly determines that $G$ has no vertex cover of size $k$ or produces a kernel $G'$ with $\Oh{k^2}$ vertices. In this case, the vertex set of $G'$ is also a vertex cover for $G$ (Proposition~\ref{prop:vc_para-l}).
	
	Running \algo{ReduceLog} on $(G, k)$ with access to $X = \V{G'}$ ($r = \abs{X} = \Oh{k^2}$) thus produces a kernel of size $r^{c_{\Pi}} (r + p(r)) = \Oh{k^{2 c_{\Pi}} (k^2 + p(k^2))}$ for $\pIndPiFDS$. Oracle access to $X = \V{G'}$ can be be provided using space $\Oh{\log{n}}$ and $\algo{ReduceLog}$ uses space $\Oh{\log{n}}$, so the total space used is $\Oh{\log{n}}$. By Theorem~\ref{lemm:para-l_kern}, any problem that can be kernelized in space $\Oh{\log{n}}$ is in $\cParaL$, and thus the claim is true.
\end{proof}

The following corollary to Theorem~\ref{thm:kern_pvc} shows how various deletion problems are in $\cParaL$ via formulations as $\pIndPiFDS$ for suitable $\Pi$.
\begin{corollary}\label{corr:vc_pidel}\exS{$\dagger$}
	Under the vertex-cover parameterization, $\pPlan$, $\pOCT$ and $\pChVD$ are in $\cParaL$. 
\end{corollary}

Via formulations as the alternative intermediate problems $\pLIndPiS$ and $\pqPartDisjPiFree$ (defined in Appendix~\ref{apdx:vc_param}), the same \algoi{ReduceLog}{$G, X, l, c_{\Pi}$} procedure with different values for $l$ can also be used to prove the next result.
\begin{corollary}\label{corr:vc_rest_gen}\exS{$\dagger$}
	Under the vertex-cover parameterization, $\pFVS$, $\pLongPath$, $\pLongCycle$ and $\pdCol$ are in $\cParaL$.
\end{corollary} \section{Solving $\pFVS$ in $5^k \cdot n^{\Oh{1}}$ time and $\Oh{k \cdot \log{n}}$ space}\label{sect:fvs}
It is known (Lemma~\ref{naivespace}) that problems in $\cFPT$ can be solved in space $f(k) + n^{\Oh{1}}$. If the sole objective is space efficiency, many problems can also be solved (na\"{i}vely, by running over all subsets of size at most $k$, for example) in $f(k) \cdot \log{n}$ bits of space, which puts them in $\cXL$. It is pertinent to note that $\cXL$ also contains problems like $\pIS$ and $\pClq$ which are $\probc{W}[1]$-complete, and hence are unlikely to be in $\cFPT$. On the other hand, it is believed that not all problems in $\cFPT$ can be solved in $f(k) \cdot \log{n}$ bits of space~\cite{CCDF1997AnnPureApplLogic}.

A third class, which requires problems in it to be simultaneously solvable in time $f(k) \cdot n^{\Oh{1}}$ and $g(k) \cdot \log{n}$ bits of space also exists, and contains $\pFVS$~\cite{EST2015Algorithmica} and $\pMOdS$~\cite{ABB+2019FSTTCS}. This class, which we call $\cFPTXL$, has been previously identified in the literature as $D[f \oper{poly}, f \oper{log}]$~\cite{EST2015Algorithmica}.

In what follows, we devise an $\cFPTXL$ algorithm for $\pFVS$ which runs in time $5^k \cdot n^{\Oh{1}}$ and uses $\Oh{k \log{n}}$ bits of space. This improves on an earlier $((2k)^k \cdot n^{\Oh{1}})$-time,  $\Oh{k \log{n}}$-space algorithm of Elberfeld et al.~\cite{EST2015Algorithmica}. Along the way, we also devise a much simpler $(3k)^k \cdot n^{\Oh{1}}$-time $\cFPTXL$ algorithm for the problem.

Known $\cFPT$ algorithms for $\pFVS$~\cite{RSS2006TALG,CFK+2015book} are based on reduction rules, one of which is the following  ``short circuiting'' rule for vertices of degree two.

\subsubsection{Degree--$2$ Reduction Rule} If there is a vertex $x$ of degree $2$ with neighbours $y$ and $z$, delete $x$ and add an edge between $y$ and $z$, while retaining pre-existing edges.

Once the rule is applied, the graph possibly has parallel edges, so we assume throughout this section that we work with multigraphs. Additionally, we can assume that the graph has no isolated or degree-$1$ vertices (as they can be safely removed), no vertices with self-loops (as they must be included in the solution and then removed) and no more than two edges between a pair of vertices (as other multi-edges do not influence the solution can be removed). The correctness of these assumptions and the rule is easy to see~\cite{CFK+2015book}. 

We first discuss the space-efficient application of these rules. The degree of a vertex can be found from the adjacency list, and so it is easy to `discard' vertices with degree at most one.
Let us remark that our algorithms (in the next subsection) deletes some (up to $k$) vertices and recursively applies these rules. Those deleted vertices are stored in a separate array using $\Oh{k \log n}$ bits, and so when checking for the degree of a vertex, we also check that array appropriately besides checking the read-only adjacency list of the graph.
one

\begin{proposition}[Elberfeld et al.~\cite{EST2015Algorithmica}, Theorem 4.13]\label{degreetwoimpl}
    On a multigraph with $n$ vertices, the Degree--$2$ rule can be applied using $\Oh{\log{n}}$ bits of additional space.
\end{proposition}

One can combine this proposition with a simple branching strategy to show that $\pFVS$ is in $\cFPTXL$.

\begin{theorem}\exS{$\dagger$}
    $\pFVS$ can be solved in time ${(3k)}^k \cdot n^{\Oh{1}}$ using $\Oh{k \log{n}}$ bits of space.
\end{theorem}

\subsection{Improved algorithm based on Iterative Compression}
We now describe a restricted-space implementation of an algorithm of Chen et al.~\cite{CFL+2008JCSS} which allows us to prove the following result.
\begin{theorem}~\label{iteratefvs}
    $\pFVS$ can be solved using $O(k \log{n})$ bits of space and in time $5^k poly(n)$.
\end{theorem}

The algorithm goes through a two stage process on an instance $(G, k)$.

\textbf{Iterative Stage} In the iterative stage, the algorithm starts with an arbitrary induced subgraph $H_0$ of $G$ with $k + 1$ vertices. The vertex set of this graph is trivially also a feedback vertex set for $H_0$. The algorithm uses a \emph{compression} algorithm which given a subgraph $H$ of $G$, and a feedback vertex set $C$ for $H$ of size $k+1$, either produces a feedback vertex set $C'$ for $H$ of size at most $k$ or determines correctly that such a feedback vertex set for $H$ does not exist. In the latter case, we correctly conclude that even $G$ has no feedback vertex set of size $k$. In the former case, both the subgraph and the (compressed) solution are included an additional vertex from the graph, and the compression step is applied again. This process continues until we cover the entire graph. During the iteration, the $(k+1)$-sized solution is stored using $O(k \log n)$ bits of space.

\textbf{Compression Stage} The compression algorithm is given a feedback vertex set $S$ of size $k+1$ for the whole graph $G$, and the goal is to determine whether it has a feedback vertex set of size at most $k$. It starts by checking for every subset $Z$ of $S$ of size at most $k$ whether there is a feedback vertex set of size at most $k$ containing $Z$. This requires storing the set $Z$ which uses $\Oh{k \log{n}}$ bits of space, and requires that $G[P]$ is a forest where $P=S\setminus Z$ which can be checked using $\Oh{\log{n}}$ bits of space (by making $\Oh{n^2}$ connectivity queries on $G[P]$). Recall that $G[V\setminus S]$ is a forest as $S$ is a feedback vertex set.

Next, as a processing step, we include all vertices $v$ of $V\setminus S$ into the solution, if $v$ has at least two neighbors in a connected component of $G[P]$ which can be checked in $\Oh{\log{n}}$ bits of space using Reingold's connectivity algorithm~\cite{Rei2008JACM}. Then for any leaf vertex $v$ in $G[V\setminus S]$, we simply branch by picking $v$ into the solution or excluding $v$ from the solution which involves including $v$ into $P$. If either branch returns \NO{}, we return \NO{}. The depth of the recursion is at most $k$ and each level uses $\Oh{\log{n}}$ bits of space, so the overall process uses $\Oh{k \log{n}}$ bits of space.

Observe that both the iterative and compression stages use $\Oh{k \log{n}}$ bits of space overall. Appendix~\ref{apdx:fvs_compress} gives pseudocode (\algo{FVSCompress}, Procedure~\ref{proc:CompressFVS}) for the compression procedure described above. For proof of correctness, we refer the reader to Chen et al.~\cite{CFL+2008JCSS}, who also prove a running time bound of $5^k \cdot n^{\Oh{1}}$ which readily carries over to this setting. We now have the following lemma.

\begin{lemma}
    \algo{FVSCompress} solves instances $(G, S, k)$ using $O(k \log{n})$ bits in time $5^k poly(n)$.
\end{lemma}

The compression procedure is called at most $n - k$ times (to include one additional vertex each time starting with the first $(k + 1)$-sized subgraph), and each time it drops a vertex and adds a new vertex, with at most $k+1$ vertices at any point of time. Thus, the iterative stage adds only an overhead of $n^{\Oh{1}}$ to the running time of \algo{FVSCompress}, and Theorem~\ref{iteratefvs} follows.

The routine either stops at an intermediate stage and declares that $(G, k)$ is a \NO{} instance or continues until $i = n$, at which stage it returns the output $T_n$ of the compression routine as a feedback vertex set for $G$ of size at most $k$.

\bibliography{external/references}
\appendix
\section{Preliminaries}
\subsection*{Proof of Lemma~\ref{lemm:para-l_kern}}
\begin{proof}
	To see that the first statement holds, consider an algorithm that solves $(A, t)$ in space $f(k) + c \log{n}$. 
\begin{enumerate}
		\item Simulate the algorithm $A$ on $I$ and use a counter to count the auxiliary space used until the algorithm $A$ uses up to $(c + 1) \log{n}$ bits of memory. If it finishes executing without using more memory, output a constant-sized instance indicating the answer (\YES{} or \NO{}). This uses space $(c+1) \log(n) + O(\log \log n)$ which is $\Oh{\log{n}}$. The second term is the space used for the counter.
		\item Otherwise, i.e.\ if the algorithm needs to use more memory, output the input instance. In this case, $f(k) + c \log{n} > (c + 1)  \log{n}$, i.e.\ $f(k) > \log{n}$, and 
$n < 2^{f(k)}$.
	\end{enumerate}
	In both cases, the output instance is a kernel.

	To see that the second statement holds, observe that once a kernel of size $g(k)$ has been obtained in space $f(k) + \Oh{\log{n}}$, it can be solved by brute force in space $\Oh{h(k)}$, where $h: \N \to \N$ is a computable function. Thus, the problem can be solved in space $f(k) + h(k) + \Oh{\log{n}}$.
\end{proof}	

\section{Hitting Set and Graph Deletion Problems}\label{apdx:deletion}
\subsection*{Proof of Lemma~\ref{lemm:hs_corr}}
\begin{proof}
Let $\mathcal {F}$ be the family of sets after Reduction Rule $2$ and $\mathcal {F'}$ be the family of sets after Reduction Rule $3$. We claim that $\mathcal{F}$ has a $k$-sized hitting set if and only if $\mathcal{F'}$ has a $k-$ sized hitting set.

If $\mathcal {F'}$ has a $k$-sized hitting set, then the same hitting set will serve for $\mathcal {F}$ as every set in $\mathcal{F'}$ is a subset of a set in $\mathcal{F}$.
To show the converse, suppose $\mathcal{F}$ has a hitting set $S$ containing an element $x$ which was not present in $\mathcal{F'}$. Suppose $x$ intersects a set $F \in \mathcal{F}$.
As $x$ was not present in $\mathcal{F'}$, then $x$ was present only in $F \in \mathcal{F}$ and hence reduction rule $3$ deleted it. But then there must be some other element $y$ that was present in $F$ and was present only in $F \in \mathcal{F}$.
Thus $S \setminus {x } \cup {y} $ will be a set that will intersect $F \in \mathcal{F}$ and is of the same size as $S$. Repeating this process for every set in $\mathcal{F'}$ not intersected by $S$, will result in a hitting set of $\mathcal{F'}$ of the same size as $S$.

Observe that after Reduction Rule $2$, the number of remaining sets is at most $k^2$. Let $F$ be any set in the family after Reduction Rule $3$, and let $x$ be any element of $F$. Then $x$ appears in only $F$ in the family (and if so, it is the only such element), or it appears in some other set of the family. But as any pair of sets has intersection at most one, if $x$ appears in some other set $F'$ of the family, no other element of $F$ can appear in $F'$. Hence every distinct element of $F$ can appear in a distinct element of $F$, and so $F$ has at most $k^2$ elements, and the lemma follows.
\end{proof}

\subsection*{Space-Efficient Implementation of Rules in Subsection~\ref{ssct:bd_hs_dg1}}
Ultimately, we need to determine the sets that exist in the family at the end of Reduction Rule $(d+1)$ and the elements that survive at the end of Reduction Rule $(d+2)$, as these are the sets and elements that will be written to the output stream. We call the input family of sets Stream $0$ and for each $i \in [d + 2]$, we call the family of sets that exists after Rule $i$ has been applied as Stream $i$.

Now a set $X$ is in Stream $i$ if and only if any of the following conditions is satisfied:
\begin{enumerate}
    \item $X$ is in Stream $(i - 1)$ and no $(d - 1 + i)$-subset of $X$ appears more than $k^i$ times in Stream $(i-1)$ or
    \item $X$ is a $(d-i+1)$ subset of the universe that appears more than $k^i$ times in Stream $(i-1)$.
\end{enumerate}

The correctness follows from the correctness of the reduction rules.
Since we don't have space to store any specific stream, we recursively check for sets in the previous stream.

If $s_i$ is the space used for checking for a set in Stream $i$, we observe that $s_i$ satisfies $s_i = \Oh{\log{n}} + (d - 1 + i) \log{n} + s_{i - 1}$. The second term is to enumerate all subsets of size $(d-1+i)$.  Since the input can be read using space $s_0 = \Oh{\log{n}}$, this recurrence yields $s_d = \Oh{d^2 \log{n}}$. Thus, a set in Stream $d$ can be checked using $\Oh{d^2 \log{n}}$ bits of space. Observe that given access to Stream $d$, Rule $d + 1$ can be applied by simply counting the number of sets in Stream $d$, which uses $\Oh{\log{n}}$ bits of additional space.

Also all sets in Stream $i$ can be generated, given access to Stream $i-1$ as follows.

\begin{itemize}
\item
For each set $X$ in Stream $i-1$
\begin{itemize}
           \item For each subset $Y$ of $X$ of size $(d-1+i)$,
           \begin{itemize}
           \item Count the number of times $Y$ appears in Stream $(i - 1)$.
           \item If this number is at least $k^i + 1$, then break out of this enumeration and move to the next set (and $X$ is not output).
           \end{itemize}
        \item Write $X$ to the output stream (if no $Y$ appears in $k^i +1$ sets of Stream $i-1$).
\end{itemize}
\item
For each $(d-1+i)$-subset $Y$ of the universe,
     \begin{itemize}
           \item Count the number of times $Y$ appears in Stream $(i - 1)$.
            \item If this number is at least $k^i + 1$, then output $Y$.
     \end{itemize}
\end{itemize}

Now Stream $(d + 2)$ which is the output, can be generated as follows.
\begin{itemize}
    \item For each set $X$ in Stream $(d + 1)$:
    \begin{itemize}
        \item Set $f_X \gets 0$.
        \item For each element $e$ of $X$:
        \begin{itemize}
            \item Count the number of times $e$ appears in Stream $(d + 1)$.
            \item If this number is at least $2$, write $e$ to the output stream and skip to the next element in this enumeration.
            \item If $f_X = 1$, skip to the next element in this enumeration.
            \item Write $e$ to the output stream and set $f_X \gets 1$.
        \end{itemize}
        \item Write a set separator to the output stream.
     \end{itemize}
\end{itemize}

The above procedure uses a constant number of extra counters to count/enumerate sets and elements in Stream $(d + 1)$, which only requires an additional $\Oh{\log{n}}$ bits of space. Thus Stream $(d + 2)$ can be generated using space $\Oh{d^2 \log{n}}$ overall, and we have the following result.

\subsection*{Algorithm for the Proof of Theorem~\ref{thm:restr_para_l}}

\subsection*{Proof of Lemma~\ref{lem:correctness}}
\begin{proof}
	Let $(G, k)$ be an instance of $\pDelPi$. Consider the family $\mathcal{F} = \setb{S \subseteq V}{\exists\ \text{injective homomorphism}\ H \to G[S]\ \text{for some}\ H \in \Phi}$. Any solution $R$ for $(G, k)$ can be decomposed as $R = S \sqcup T$, where $S$ is a minimal set of vertices such that $G - S$ is in $\Psi$. This follows directly from the fact that $\Pi \subseteq \Psi$. The family $\mathcal{F}$ consists of precisely those sets of vertices in $G$ which must be hit by $S$ so that $G - S \in \Psi$. Thus, $S$ is a minimal hitting set for $\mathcal{F}$ if and only if it is a minimal set of vertices such that $G - S \in \Pi$.

	Algorithm~\ref{algo:SolveDelPi} first constructs the \pdHS{} instance $(V, \mathcal{F}, k)$ and then using \algo{KernelizeDeeHS}, it obtains the kernel $(V', \mathcal{F}', k)$. Then by the branching procedure \algo{BranchAndCall}, it attempts to find a minimal set $S$ of size at most $k$ such that $G - S \in \Psi$.

	Recall that any minimal set of size at most $k$ which hits all of $\mathcal{F}'$ also a hits all of $\mathcal{F}$ and vice versa (Proposition~\ref{prop:dhs_log_kern}). In the \algo{BranchAndCall} procedure, the algorithm branches on sets appearing in $\mathcal{F}'$. At each recursive call, one of the at most $d$ vertices in the set is added to $S$.

	In this way, all possible minimal hitting sets of size at most $k$ (if they exist) for $\mathcal{F}'$ and therefore also $\mathcal{F}$ are explored. At the leaves of the branching tree where $l \geq 0$, a minimal hitting set $S$ for $\mathcal{F}$ of size at most $k - l$ is obtained.

	In the other case, i.e.\ when $l < 0$ at a leaf of the branching tree, the sequence of vertices picked on the branching path cannot yield a set $S$ of size at most $k$ such that $G - S \in \Psi$. If this condition occurs at all the leaves of the branching tree, the procedure correctly indicates that the input is a \NO{} instance.

	Now at the leaves of the branching tree where $l \geq 0$, the algorithm executes \algoi{SolveDel\textPi{}On\textPsi{}}{$G - S, l$}, which checks if  at most $l$ more vertices can be deleted from $G - S$ such that the resulting graph is in $\Psi$. Observe that this occurs for every minimal set $S$ of size $k$ such that $G - S \in \Psi$. Thus, if the output of \algoi{SolveDel\textPi{}On\textPsi{}}{$G - S, l$} on any of these sets $S$ is \YES{}, then the input is a \YES{} instance of $\pDelPi$. Because all such minimal sets are explored in the branching tree, if the output of \algoi{SolveDel\textPi{}On\textPsi{}}{$G - S, l$} at each of the leaves is \NO{}, then the algorithm correctly determines that the input is a \NO{} instance. 
\end{proof}

\subsection*{Proof of Theorem~\ref{thm:restr_para_l}}
\begin{proof}
	Observe that $(V, \mathcal{F}, k)$ can be generated implicitly by enumerating all sets $S \subseteq V$ of sizes $c_1, \dotsc, c_t$, and only retaining those sets $S$ for which a graph in $\Phi$ has an injective homomorphism to $G[S]$. Since the graphs in $\Phi$ are of constant size and there are finitely many of them, this can be done in space $\Oh{\log{n}}$. 

	The initial kernelization step uses space $\Oh{\log{n}}$ (Proposition~\ref{prop:dhs_log_kern}). The later branching steps have implicit access to the kernel, and at each branch, the vertex $v \in \mathcal V' \subseteq V$ to be deleted must be stored in memory. Performing this deletion na\"{i}vely would require space $\Oh{\log{n}}$. However, the vertices to be deleted are picked from $V'$, which has $\Oh{(k + 1)^d}$ elements (Proposition~\ref{prop:dhs_log_kern}). Thus, the vertices deleted in the branching steps can be stored in memory as their indices in the vertex set $V'$, which uses $\Oh{\log{k}}$ space per vertex. Thus, all the vertices deleted on any branching path can be stored using space $\Oh{k \log{k}}$.

	At every leaf of the branching tree, the graph $G - S$ can be represented implicitly using space $\Oh{\log{n}}$ because $G - S = G[V \setminus S]$ is an induced subgraph of $G$. The subsequent call to \algo{SolveDel\textPi{}On\textPsi{}} uses space $f(k) + \log{n}$, so the total space used by the procedure is $\Oh{h(k) + k \log{k} + \log{n}} = g(k) + \Oh{\log{n}}$ for some computable function $g: \N \to \N$. This proves the claim.
\end{proof}

\subsection*{Problem Definitions}\label{apdx:del_problems}
The problems considered in Section~\ref{sect:deletion_problems} section are defined as follows.
\begin{description}
	\item[\pDLF{}]\hfill\\
		\textbf{Instance:} $(G, k)$, where $G$ is a graph and $k \in \N$\\
		\textbf{Question:} Is there a set $S \subseteq V$ with $\abs{S} \leq k$ such that $G - S$ is a linear forest, i.e.\ a forest consisting exclusively of vertex-disjoint paths? 
	\item[\pDPo{}]\hfill\\
		\textbf{Instance:} $(G, k)$, where $G$ is a graph and $k \in \N$\\
		\textbf{Question:} Is there a set $S \subseteq V$ with $\abs{S} \leq k$ such that $G - S$ is a graph of pathwidth $1$? 
\end{description}

\begin{proposition}[Reingold~\cite{Rei2008JACM}, Theorem 4.1]\label{prop:connectivity}
	In any graph on $n$ vertices, it can be determined if any two vertices are connected by a path using space $\Oh{\log{n}}$.
\end{proposition}

\begin{proposition}[Philip et al.~\cite{PRV2010WG}, Lemma 3]\label{prop:dpo_forb}
	A graph which is $T_2$-, $C_3$- and $C_4$-free is a disjoint union of trees, cycles and cycles with some pendant vertices.
\end{proposition}

\subsection*{Proof of Corollary~\ref{corr:dlf_pw1}}
\begin{proof}
	\begin{claim}
		$\pDLF{} \in \cParaL$.
	\end{claim}
	Observe that in a linear forest, no vertex can have more than $2$ neighbours, i.e.\ the forest does not include $T_1$ as a subgraph. Conversely, any graph that does not include $T_1$ as a subgraph can only have vertices of degree at most $2$, i.e.\ it consists exclusively of paths and cycles. In that case, deleting exactly one vertex from each cycle yields a linear forest, i.e. the number of vertices that must be deleted is equal to the number of connected components that are not trees. Procedure~\ref{proc:SolveDLFPOnPsi} shows how this can be done. The correctness of the procedure is immediate from the preceding discussion.

	To see that it uses space $\Oh{\log{n}}$, note that using the algorithm of Proposition~\ref{prop:connectivity}, the connected component of any vertex can be determined in space $\Oh{\log{n}}$. To determine if $G[C]$ is not a tree, it suffices to count the number of edges in G[C], which exceeds $\abs{C} - 1$ if and only if $G[C]$ is not a tree. With access to $C$, this can be done in space $\Oh{\log{n}}$. All other steps in the procedure also use space $\Oh{\log{n}}$, which makes a total of $\Oh{\log{n}}$ for the entire procedure.

	Thus, $\pDLF$ restricted to $\Psi^2$, the class of graphs of degree at most $2$ is in $\cL{} \subseteq \cParaL$. Since $T_1$ characterizes $\Psi^2$ as a forbidden induced subgraph, the claim follows from Theorem~\ref{thm:restr_para_l}.

	\begin{claim}
		$\pDPo{} \in \cParaL$.
	\end{claim}
	Consider the class of graphs $\Psi_3$ which do not include $T_2, C_3$ or $C_4$ as a subgraph. Any graph of pathwidth at most $1$ cannot include a $T_2$ or a cycle, so $\Psi_3$ includes all graphs of pathwidth at most $1$. By Proposition~\ref{prop:dpo_forb}, graphs in $\Psi_3$ are disjoint unions of trees, cycles and cycles with pendant vertices. To convert such a graph to a graph of pathwidth at most $1$, exactly one vertex from each cycle must be deleted. Equivalently, the number of vertices to be deleted is equal to the number of connected components that are not trees. Consider Procedure~\ref{proc:SolveDLFPOnPsi}. From the preceding discussion, it is clear that the procedure also correctly solves $\pDPo$ on $\Psi_3$. The space used by the procedure, as before, is $\Oh{\log{n}}$.
\end{proof}

\begin{procedure}

	$l \gets 0$\;
	\For{$v \in V$}{
		determine $C$, the set of vertices in the connected component of $v$\;
		\If{$v$ is the smallest vertex in $C$ and $G[C]$ is not a tree}{
			$l \gets l + 1$\;
			\If{$l > k$}{
				\Return{\NO}\;
			}
		}
	}
	\Return{\YES}\;

\caption{SolveDLFP1On\textPsi{}(): solve a restricted graph deletion problem}\label{proc:SolveDLFPOnPsi}
\end{procedure}

\section{Vertex Cover Number as Parameter}\label{apdx:vc_param}

\subsection*{The \algo{ReduceLog} Procedure}
\begin{procedure}
	\KwIn{$(G, X, l, c_{\Pi})$, where $G$ is a graph, $X$ is a vertex cover for $G$, and $l, c_{\Pi} \in \N$}
	\KwOut{$G'$, a reduced graph}
	
		$S \gets \emptyset$\;
		\For{$v \in \V{G}$}{\label{proc:ReduceLog:outer_loop}
			\For{$Y \subseteq X$ with $\abs{Y} \leq c_{\Pi}$}{\label{proc:ReduceLog:main_loop}
				\For{each partition $Y = Y^+ \sqcup Y^-$}{\label{proc:ReduceLog:inner_loop}
					$Z \gets \setb{w \in \V{G} \setminus X}{w\ \text{is adjacent to all of}\ Y^+\ \text{and none of}\ Y^-}$\;
					\If{$v$ is among the first $l$ vertices in $Z$}{\label{proc:ReduceLog:check}
						$S \gets S \cup \brc{v}$\;\label{proc:ReduceLog:augment}
					}
				}
			}
		}
		\Write{$G - S$}\;
	
	\caption{ReduceLog(): reduce $G$ using the vertex cover $X$}\label{proc:ReduceLog}
\end{procedure}

The orginal \algo{Reduce} procedure, described below, differs only in the way the vertices $v \in Z$ to be added to $S$ are determined. 

\begin{procedure}
\KwIn{$(G, X, l, c_{\Pi})$, where $G$ is a graph on $n$ vertices, $X$ is a vertex cover for $G$, and $l, c_{\Pi} \in \N$}
\KwOut{$G'$, a reduced graph}

	$S \gets \emptyset$\;
	\For{$Y \subseteq X$ with $\abs{Y} \leq c_{\Pi}$}{\label{proc:Reduce:main_loop}
		\For{each partition $Y = Y^+ \sqcup Y^-$}{
			let $Z$ be the set of vertices in $\V{G} \setminus X$ adjacent to all of $Y^+$ and none of $Y^-$\;
			let $T$ be an arbitrary set of $l$ vertices of $Z$ or all of it, if $Z$ has less than $l$ vertices\;
			$S \gets S \cup T$;
		}
		\Write{$G - S$}\;
	}

\caption{Reduce(): reduce $G$ using the vertex cover $X$}\label{proc:Reduce}
\end{procedure}

The original \algo{Reduce} procedure marks vertices $v \in G$ if $v$ is adjacent to all of $Y^+$ and none of $Y^-$ for some partition $Y^+ \sqcup Y^-$ of $X$. This may require space $\Om{n}$. In contrast, \algo{ReduceLog} bypasses this marking step while producing the same kernelization guarantees as in Proposition~\ref{prop:reduce_pvc_kernels}. Correctness is easy to see by comparing the procedures \algo{ReduceLog} and \algo{Reduce}. We now prove Lemma~\ref{lemm:reduce_log_pvc}.

\begin{proof}
	The space used by the loop at Line~\ref{proc:ReduceLog:outer_loop} is $\Oh{\log{n}}$, since it only requires the vertex $v \in \V{G}$ to be stored. The loop at Line~\ref{proc:ReduceLog:main_loop} must store a set $Y \subseteq X$ of size at most $c_{\Pi}$. With access to $X$, individual elements within it can be referenced using $\Oh{\log{\abs{X}}}$ bits of space, so $Y$ can be stored using space $\Oh{c_{\Pi} \log{\abs{X}}}$. Similarly, the partition $Y^+ \sqcup Y^- = Y$ in the loop at Line~\ref{proc:ReduceLog:inner_loop} can be stored using space $\Oh{c_{\Pi} \log{\abs{X}}}$.

	Now the set $Z$ can be determined by enumeration: for each $v \in \V{G}$, enumerate $Y^+$ and check if $v$ is adjacent to all vertices enumerated, and then enumerate $Y^-$ and check if $v$ is adjacent to none of the vertices enumerated. This uses space $\Oh{\log{n}}$ (to enumerate all vertices $v \in \V{G}$), in addition to the space used to store $Y^+$ and $Y^-$.

	Observe that the set $S$ is only modified on Line~\ref{proc:ReduceLog:augment}, and in fact, the loop at Line~\ref{proc:ReduceLog:outer_loop} simply determines for every vertex $v \in \V{G}$ whether $v$ is in $S$. Thus, $S$ can be determined implicitly using the loop; the check at Line~\ref{proc:ReduceLog:check} determines whether $v$ is in $S$. The output of the procedure is the induced graph $G - S$, which, given access to $S$, can be produced by enumerating the vertices and edges of $G$ and suppressing vertices which appear in $S$ (this can again be checked by enumerating $S$) and edges whose endpoints appear in $S$. This uses an additional $\Oh{\log{n}}$ bits of space. Thus, the total space used by the procedure is $\Oh{c_{\Pi} \log{\abs{X}} + \log{n}} = \Oh{\log{n}}$, since $\abs{X} \leq n$, and $c_{\Pi}$ is a constant.
\end{proof}

\subsection*{Problem Definitions}
\begin{description}
	\item[$\pLIndPiS$]\hfill\\
		$\Pi$ is a class of possibly empty graphs and there is a non-decreasing polynomial $p$ such that every $G \in \Pi$ satisfies $\abs{G} \leq p(\tau_G)$.\\
		\textbf{Instance:} $(G, k)$, where $G$ is a graph and $k \in \N$\\
		\textbf{Question:} Is there a set $S \subseteq V$ with $\abs{S} \geq k$ such that $G[S] \in \Pi$?
	\item[$\pqPartDisjPiFree$]\hfill\\
		$\Pi$ is class of possibly empty graphs and there is a non-decreasing polynomial $p$ such that every vertex minimal $G \in \Pi$ satisfies $\abs{G} \leq p(\tau_G)$.\\
		\textbf{Instance:} $(G, k)$, where $G$ is a graph and $k \in \N$\\
		\textbf{Question:} Is there a partition $\V{G} = V_1 \sqcup \dotsc \sqcup V_q$ such that $G[V_i]\ (i \in [q])$ contains no graph in $\Pi$ as an induced subgraph?
	\item[$\pPlan$]\hfill\\
		\textbf{Instance:} $(G, k)$, where $G$ is a graph and $k \in \N$\\
		\textbf{Question:} Is there a set $S \subseteq V$ with $\abs{S} \leq k$ such that $G - S$ is a planar graph?
	\item[$\pOCT$]\hfill\\
		\textbf{Instance:} $(G, k)$, where $G$ is a graph and $k \in \N$\\
		\textbf{Question:} Is there a set $S \subseteq V$ with $\abs{S} \leq k$ such that $G - S$ is a bipartite graph?
	\item[$\pChVD$]\hfill\\
		\textbf{Instance:} $(G, k)$, where $G$ is a graph and $k \in \N$\\
		\textbf{Question:} Is there a set $S \subseteq V$ with $\abs{S} \leq k$ such that $G - S$ is a chordal graph?
	\item[$\pFVS$]\hfill\\
		\textbf{Instance:} $(G, k)$, where $G$ is a graph and $k \in \N$\\
		\textbf{Question:} Is there a set $S \subseteq V$ with $\abs{S} \leq k$ such that $G - S$ is a forest?
	\item[$\prob{Longest Path/Cycle}$]\hfill\\
		\textbf{Instance:} $(G, k)$, where $G$ is a graph and $k \in \N$\\
		\textbf{Question:} Does $G$ contain a path/cycle of length at least $k$?
\end{description}

\subsection*{Formulating problems as $\pIndPiFDS$, $\pLIndPiS$ or $\pqPartDisjPiFree$}
\begin{description}
	\item[$\pPlan$]\hfill
		\begin{itemize}
			\item forbidden class is the class of graphs containing $K_{3, 3}$ or $K_5$ minors, characterized by $c_{\Pi} = 4$ adjacencies
			\item equivalent to $\pIndPiFDS$
		\end{itemize}
	\item[$\pOCT$]\hfill
		\begin{itemize}
			\item forbidden class is the class of odd-length cycles characterized by $c_{\Pi} = 2$ adjacencies
			\item equivalent to $\pIndPiFDS$
		\end{itemize}
	\item[$\pChVD$]\hfill
		\begin{itemize}
			\item forbidden class is the class of chordless cycles characterized by $c_{\Pi} = 3$ adjacencies
			\item equivalent to $\pIndPiFDS$
		\end{itemize}
	\item[$\pFVS$]\hfill
		\begin{itemize}
			\item forbidden class is the class of cycles characterized by $c_{\Pi} = 2$ adjacencies
			\item equivalent to $\pqPartDisjPiFree$
		\end{itemize}
	\item[$\pLongPath$]\hfill
		\begin{itemize}
			\item target class is the class of hamiltonian paths characterized by $c_{\Pi} = 2$ adjacencies
			\item equivalent to $\pLIndPiS$
		\end{itemize}
	\item[$\pLongCycle$]\hfill
		\begin{itemize}
			\item target class is the class of hamiltonian cycles characterized by $c_{\Pi} = 2$ adjacencies
			\item equivalent to $\pLIndPiS$
		\end{itemize}
\end{description}

The above formulations show that all problems appearing in Corollaries~\ref{corr:vc_pidel}~and~\ref{corr:vc_rest_gen} can be formulated as $\pDelPi$, $\pLIndPiS$ or $\pqPartDisjPiFree$, from which it is clear that they can be kernelized using \algo{ReduceLog}. For further details, we refer the reader to Fomin et.\ al~\cite{FJP2014JCSS}.

\section{$\pFVS$}\label{apdx:fvs_compress}

The following procedure determines whether a pair of vertices $u$ and $v$ are adjacent \emph{after} applying the Degree-$2$ rule.

\subsection*{Solving $\pFVS$ in time ${(3k)}^k \cdot n^{\Oh{1}}$ using $\Oh{k \log{n}}$ bits of space}
The following result allows us to use a simple branching strategy to solve $\pFVS$.
\begin{proposition}[Cygan et al.~\cite{CFK+2015book}]\label{top3kdegree}
    Let $G$ be a graph with minimum degree $3$ and let $v_1, \dotsc, v_n$ be the vertices of $G$ ordered by degree from smallest to largest. Any any feedback vertex set for $G$ of size at most $k$ ($3k \leq n$) contains one of the vertices $v_1, \dotsc, v_{3k}$.
\end{proposition}

Consider the following algorithm. Start by applying the Degree--$2$ Rule and remove vertices that are isolated, have degree one and have self-loops (in this case, decrease the solution budget by one). Now branch by picking one of the $3k$ highest-degree vertices into the solution and decrease the solution budget by one. At the beginning of each branch, reapply the Degree--$2$ rule and the rules for vertices that are isolated, have degree one or have self loops. After $k$ steps of this branching process, check whether the remaining graph is a forest. If it is, output the solution vertices and if all the branches fail, return \NO{}.

The last step of checking if the graph is a forest can be performed using space $\Oh{\log{n}}$ (see the proof of Corollary~\ref{corr:dlf_pw1}) and as discussed earlier each of the reduction rules can be performed using space $\Oh{\log{n}}$. As the height of the recursion is at most $k$, the space used to store the solution vertices is $\Oh{k \log{n}}$. Thus, the algorithm uses a total of $\Oh{k \log{n}}$ bits of space. Additionally, the algorithm branches on $3k$ vertices at each level of the recursion, so the running time is ${(3k)}^k \cdot n^{\Oh{1}}$. The next theorem follows directly from this discussion.

\subsection*{Solving $\pFVS$ in time $5^k \cdot n^{\Oh{1}}$ using $\Oh{k \log{n}}$ bits of space}

\begin{procedure}[h]
\KwIn{$(G, S, k)$, where $G$ is a graph, $k \in \N$ and $S$ is a feedback vertex set for $G$ of size at most $k + 1$}
\KwOut{$S'$, a feedback vertex set for $G$ of size at most $k$ or $\NO$ if no such set exists}
    \For{each $Z \subseteq S$ of size at most $k$}{
        \Return{\algoi{FVSBranch}{$G - Z, S \setminus Z, \emptyset, k - \abs{Z}$}}\;
    }

    \SetKwFunction{brchproc}{FVSBranch}
    \SetKwProg{subproc}{Subroutine}{}{}
    \subproc{\brchproc{$H, P, R, l$}}{
        \If{$l \leq 0$}{
            return $R$ if $H$ is a forest and \NO{} otherwise\;
        }

        \For{each $v \in \V{H} \setminus R$}{
            \If{there is a cycle $C$ such that $\V{C} \setminus P = \brc{v}$}{
                $R \gets R \cup \brc{v}, l \gets l  - 1$\;
            }
        }

        apply the Degree-2 rule exhaustively to $H$\;\label{proc:CompressFVS:applydeg2}
        \For{each leaf $v \in H \setminus P$}{
            \For{all $u, w \in \Nbr{P}{v}$}{
                \If{$u$ and $w$ are connected in $H[P]$}{
                    $U \gets \algoi{FVSBranch}{H, P, R \cup \brc{v}, l - 1}, W \gets \algoi{FVSBranch}{H, P \cup \brc{v}, R, l}$\;
                    if $U \neq \NO$ then return $U$\;
                    if $W \neq \NO$ then return $W$\;
                    \Return{\NO{}}\;
                }
            }
        }
    }

    \caption{FVSCompress(): find a smaller feedback vertex set given one as input}\label{proc:CompressFVS}
\end{procedure}
 \end{document}